# Spintronic Emitters for Super-Resolution in THz-Spectral Imaging

Finn-Frederik Stiewe[1], Tristan Winkel[1], Yuta Sasaki[2,3], Tobias Tubandt[1], Tobias Kleinke[1], Christian Denker[1], Ulrike Martens[1], Nina Meyer[1], Tahereh Sadat Parvini[1], Shigemi Mizukami[3,4,5], Jakob Walowski[1*] and Markus Münzenberg[1]

[1]Institut für Physik, Universität Greifswald, Greifswald, Germany
[2]Department of Applied Physics, Graduate school of Engineering, Tohoku University, Sendai, Japan
[3]WPI Advanced Institute for Materials Research (AIMR), Tohoku University, Sendai, Japan
[4]Center for Spintronic Research Network (CSRN), Tohoku University, Sendai, Japan
[5]Center for Science and Innovation in Spintronics (CSIS), Tohoku University, Sendai, Japan

(Dated: October 28, 2021)

**Abstract.** THz-spectroscopy is an attractive imaging tool for scientific research, especially in life science, offering non-destructive interaction with matter due to its low photon energies. However, wavelengths above $100~\mu m$ principally limit its spatial resolution in the far-field by diffraction to this regime, making it not sufficient to image biological cells in the micrometer scale. Therefore, super-resolution imaging techniques are required to overcome this restriction. Near-field-imaging using spintronic emitters offers the most feasible approach because of its simplicity and potential for wide-ranging applications. In our study, we investigate THz-radiation generated by fs-laser-pulses in CoFeB/Pt heterostructures, based on spin currents, detected by commercial LT-GaAs Auston switches. The spatial resolution is evaluated applying a 2D scanning technique with motorized stages allowing scanning steps in the sub-micrometer range. By applying near-field imaging we can increase the spatial resolution to the dimensions of the laser spot size in the micrometer scale. For this purpose, the spintronic emitter is directly evaporated on a gold test pattern separated by a $300~nm$ spacer layer. Moving these structures with respect to the femtosecond laser spot which generates the THz radiation allows for resolution determination using the knife-edge method. We observe a full-width half-maximum THz beam diameter of $4.9 \pm 0.4~\mu m$ at $1~THz$. The possibility to deposit spintronic emitter heterostructures on simple glass substrates makes them an interesting candidate for near-field imaging for a large number of applications.

Terahertz ($0.6 - 10~THz$) radiation provides a great potential for biomedical applications such as biosensing for cancer imaging due to its low non-ionizing photon energies and thus a non-destructive interaction of THz radiation with tissue [1] [2] [3] [4]. In addition, THz spectroscopy determines specific spectral fingerprints for numerous materials which are not interfering with other spectral ranges and especially water-based materials show a strong absorption [5] [6]. This imaging technique is already used in airports, to distinguish macroscopically sized materials from human cell tissue. Increasing the imaging resolution to the micrometer scale will provide new insights into biological cells with diameters from $1~\mu m$ to $100~\mu m$ [7] or even the determination of impurities and intoxicants enriched inside them. Further improvement will enable the observation of genetic information (DNA) [8] [9] [10].

In accordance with Abbe´s diffraction limit, the spatial resolution of microscopy techniques is defined by half of the wavelength which for THz radiation yields to a few hundred micrometers. The simplest approach to overcome this limitation is near-field imaging [11] [12] [13]. For this purpose, the light source spot needs to be limited by an aperture and simultaneously, the THz radiation needs to be generated in direct vicinity to the investigated structure. Where both the aperture and the distance to the structure need to be much smaller than the generated wavelength. In principle, the dimensions of these two factors limit the possible resolution. Several studies have already demonstrated the huge potential of THz near-field imaging for the investigation of optical near-fields close to metallic structures, such as small apertures [14] [15] [16] [17] [18], or microresonators and metamaterials [19] [20]. In those investigations, THz pulses with large beam dimensions are generated in the far-field and propagate through apertures with subwavelength dimensions to achieve reduced beam diameters. The diffracted waves need to be detected behind those apertures at distances smaller than the wavelengths, before diffraction effects occur [21] [22] [23] [24]. Further, there is development in near field imaging by illuminating whole arrays and reconstructing the images by calculating the signal correlation [25]. In this regard, spintronic THz emitters exhibit several possibilities for implementation with biological samples in combination with near field imaging. They are low cost in production and can be mass produced by depositing the bilayer heterostructures on a huge variety of glass substrates. Besides this, they offer polarization control and therefore the choice of materials to be examined can be extended to magnetic particles e.g., in functionalized cell tissue. In this study, we investigate spintronic THz emitters which can be placed in direct proximity to the imaged structure. THz radiation is generated by femtosecond laser pulses focused down to the micrometer scale, and thus confining the spatial expansion of the THz source spot to those dimensions. The specially designed and lithographically prepared test structure consists of an adjacent patterned gold film, separated from the emitter by an insulating spacer layer. The THz pulse generation in direct vicinity of the imaged object in general permits a near-field independent detection because the imaging itself is not affected by diffraction effects. By systematically moving the sample, a THz pulse is generated, and the gold pattern is imaged. The results confirm a high near-field spatial resolution of $4.9 \pm 0.4$ μm for the frequency of $1$ THz, which is in the range of the laser spot size. With this approach, the capability for near-field imaging at high spatial resolutions overcomes the Abbe limit by 30 times [26] [27] [28].

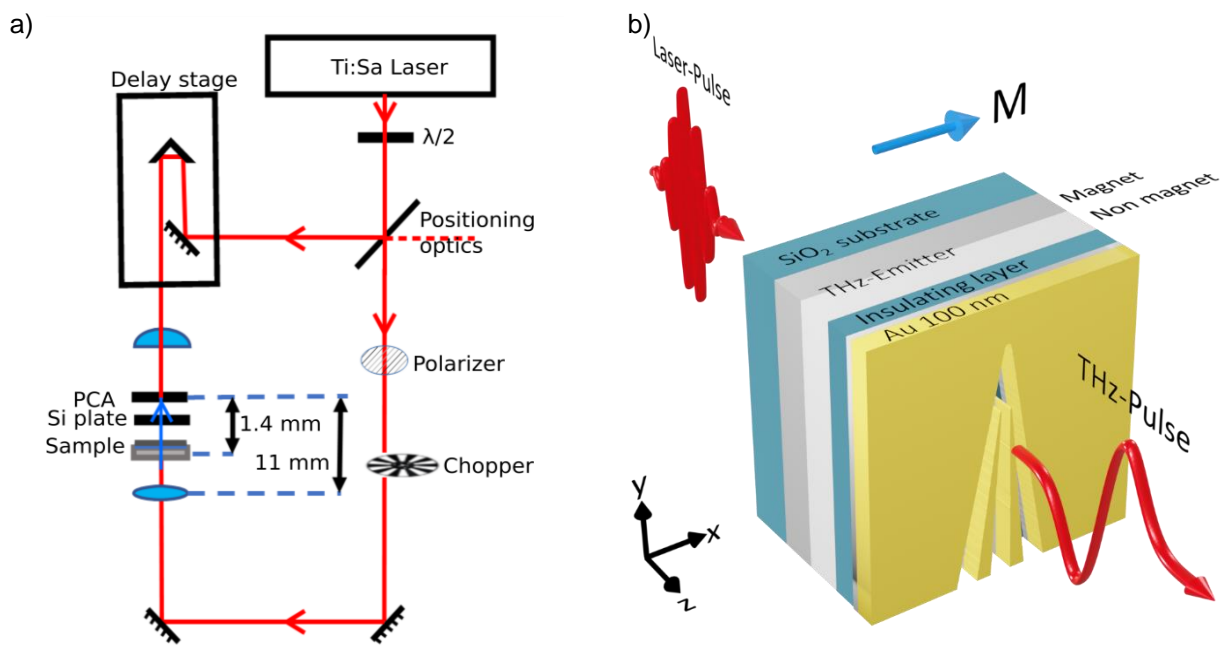

**Figure 1:** a) Scheme of the optical setup. A Ti:sapphire laser system (central wavelength $810$ nm, repetition rate $80$ MHz, pulse duration $40$ fs) is used as a femtosecond optical source. The laser is split into two parts, a pump beam and a probe beam. The pump beam is guided to the spintronic emitter (sample) which emits the THz-radiation. It is modulated by a chopper with a frequency of $1.5$ kHz to improve the signal-to-noise ratio with a lock-in amplifier. The lock-in amplifier is connected to the detector, an Auston switch (photoconductive antenna (PCA)). The lens in front of the spintronic emitter confines the pump beam to the micrometer range. The probe beam is guided through a variable path for probe beam delay adjustment and focused onto the backside of the PCA. For zero delay between the probe laser pulse exciting free charge carriers and the THz-pulse generating an electric field, the resulting electric current in the PCA yields the measurement signal. b) Schematic of sample layout together with principle of spintronic terahertz emitter. A spin current is excited by the femtosecond laser beam in the non-magnetic/ferromagnetic (NM/FM) heterostructure, and then transient charge currents are generated by the inverse spin Hall effect (ISHE), leading to THz emission out of the structure.

The schematic of the optical setup used for near-field THz imaging is shown in Figure 1a). A Ti:Sa laser (Coherent Vitara) with $40$ fs pulse duration at $810$ nm central wavelength and a repetition rate of $80$ MHz is used as an optical light source for THz pulse generation and stroboscopic detection. The laser beam is split into a pump and a probe beam. The intense pump beam is guided to the emitter to generate THz radiation. The probe beam is directed through an adjustable temporal delay, which allows for systematic THz pulse reconstruction in the detection scheme. The THz emitter is a spintronic heterostructure consisting of a ferromagnetic FM and a heavy metal non-magnetic layer NM [29] [30] [31]. Upon laser excitation in the FM hot electrons are generated and propagate towards the NM layer (z-direction). The majority electrons experience less scattering than the minority electrons and thus have longer lifetimes in the excited state [32]. This ensures, a spin-polarized current $\mathbf{J}_s$ arriving in the NM layer. The high spin polarization leads to a motion deflection perpendicular to the propagation direction caused by the inverse Spin-Hall-Effect (ISHE) of the electrons in the NM layer [29] [33] [34] [35]. After cooling down, by exchanging energy with the lattice, the electrons relax back to their original positions. The process induces a charge current pulse $\mathbf{J}_C = \Phi_{SH} \mathbf{J}_s \times \frac{\mathbf{M}}{|\mathbf{M}|}$, where $\Phi_{SH}$ is the spin-Hall angle, and $\mathbf{M}$ the magnetization. As depicted in figure 1b), $\mathbf{J}_C$ flows in y-direction and occurs on the picosecond time scale corresponding to a Hertzian dipole emitting in the THz frequency range in z-direction. This method to generate THz radiation is meanwhile well established in various applications [36] [37] [38] [39] [40].

An Auston switch (photo-conductive antenna (PCA)) is used as a detector for the THz-radiation. The femtosecond laser pulse excites electrons in the PCA gap shorting it. The THz pulse electric field arriving at the gap sets a potential for the excited electrons generating a current and thus, the detection signal. The current direction is set by the electric field oscillation.

For spin polarization enhancement, the sample is placed in a saturating magnetic field of $10\,\mathrm{mT}$. Its direction determines the phase and polarization of the THz field, which is perpendicular to the direction of the magnetic field. Thus, the THz field polarization is incorporated into the experiment and completely controllable by the magnetic field [29]. The 2D scanning capability is implemented by two motorized μm-stages, moving the sample horizontally and vertically (x- and y-direction) with a minimum stepsize of $200\,\mathrm{nm}$ perpendicular to the laser and the THz beam propagation in z-direction. The frequencies contained in the measured THz pulses are extracted using fast Fourier transformation (FFT) [41].

As schematically depicted in figure 1b), the sample used to study THz near-field imaging consists of two different units fabricated in two steps. The first unit is a spintronic emitter ($Co_{40}Fe_{40}B_{20}$ (2 nm)/ Pt (2 nm)). The second unit is a Cr (5 nm)/Au (100 nm) test pattern layer for imaging and resolution determination. Both are separated by a $SiO_2$ (300 nm) spacer layer, which is much thinner than the THz wavelengths. This layout ensures near-field imaging and at the same time electric insulation between the emitter and Au layer. In this construction a constant distance between both units throughout the whole sample provides the necessary condition for identical THz pulses and propagation path at each position. The first unit is fabricated in the first step, where the CoFeB layer is magnetron sputtered to establish a homogeneously smooth interface with the e-beam evaporated heavy metal Pt layer. For an equidistant separation of both units, the $SiO_2$ layer is prepared by e-beam evaporation within the first preparation step. The imaging pattern is created lithographically in the second preparation step and subsequent deposition of the Cr and Au layers, where Cr serves as an adhesive layer for the Au.

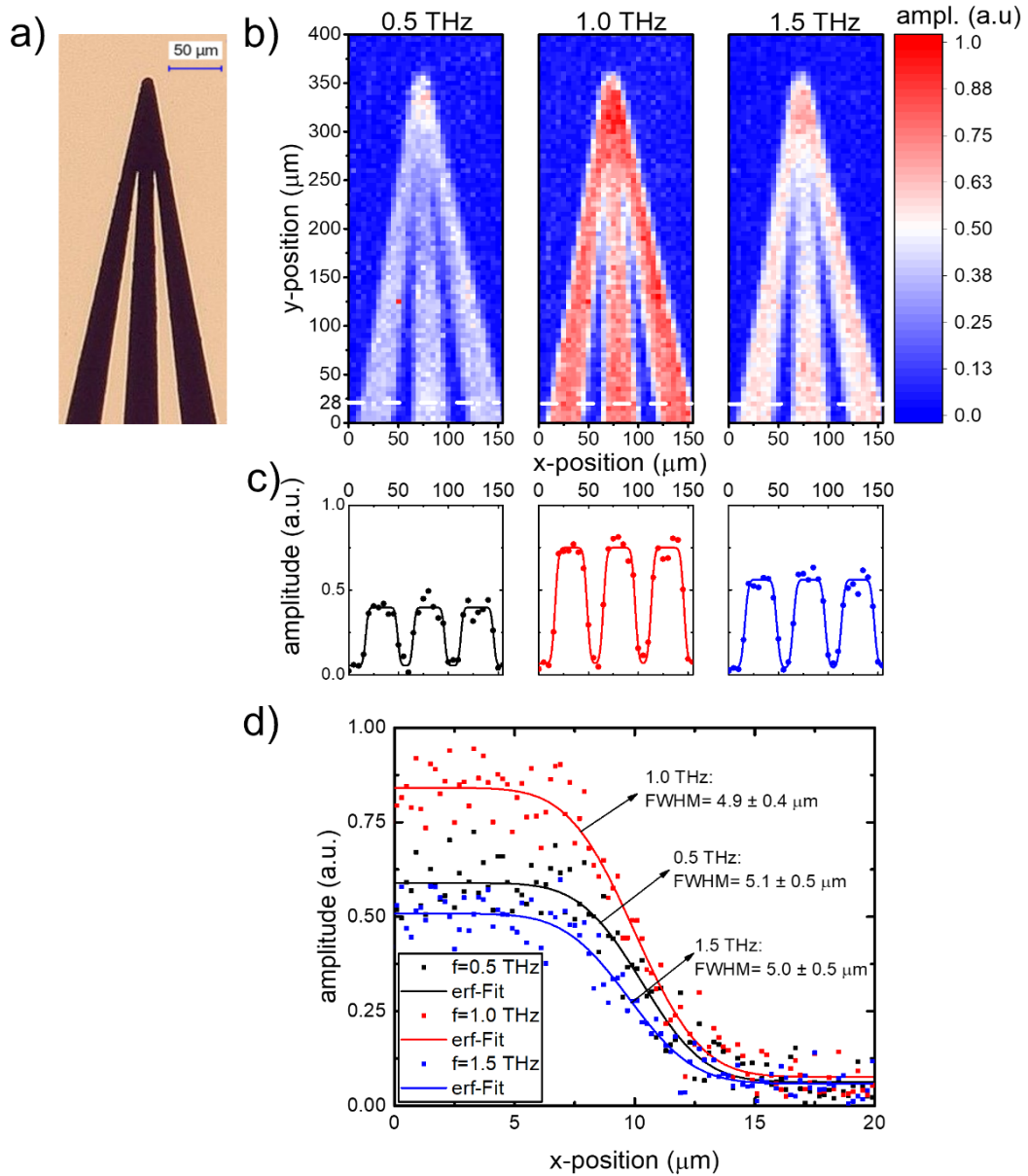

**Figure 2:** a) Optical microscope image of the designed aperture pattern. b) Two-dimensional images extracted from near-field THz scanning measurements for 0.5, 1.0 and 1.5 THz. The white dashed lines indicate the position of the intensity profiles shown in c), respectively. The error function fits (lines) to the extracted data (dots) determine the beam dimensions in the order of 5 μm. d) Additional small step size knife-edge-method data for precise resolution determination. The error function fits (erf-Fit, lines) to the amplitude profiles determine the FWHM in the plane of the gold structure on the sample for selected THz frequencies, as indicated by the arrows.

Figure 2a) depicts a light microscope image of the test pattern deposited in direct vicinity to the spintronic emitter layer. The golden area depicts the Au layer blocking the THz radiation, whereas the three slits ensuring full THz transmission are shown in brown color. This triangular inter digital pattern allows to adjust the slot size continuously up to ∼50 μm width.

Figure 2b) displays two-dimensional THz images of the test pattern in scale with figure 2a) for three extracted THz frequencies obtained by scanning the sample in x- and y-direction using a step size of 5 μm. At each xy-position, a temporal spectrum of the THz pulse is recorded. Using FFT, the included frequencies are extracted from each spectrum. The extracted frequencies are normed to the electric field amplitude $A$ of the 1 THz contribution in the signal (red), whereby, the 0.5 THz and the 1.5 THz signals (white) are generated with about half the amplitude of the 1 THz signal. For all three THz frequencies, the slit structure is clearly imaged, with the Au layer completely blocking all THz radiation (blue). In the next step, the resolution is determined using the Rayleigh criterion. According to this, considering Airy disk shape light distribution, it is possible to distinguish between two equal point sources if they are separated by a distance equal to the length between the center and the first minimum. That means, the maximum of the second point source must be located in the minimum of the first Airy disk. In this case, the amplitude in the overlap region drops down to 73.5% of both sources maximum amplitude providing a contrast of 15% and thus discriminability [42]. In another view, the full-width half-maximum (FWHM) diameter of the light source determines the lower limit for the resolution. In our approach the emitted THz pulses have a Gaussian amplitude distribution around the propagation axis, originating from the shape of the exciting laser beam. Therefore, the FWHM diameter of the emitted THz waves in the near-field limit determines the possible resolution. In figure 2c) the amplitudes extracted at the cross-section along the white lines in figure 2b) show multiple transitions between the THz radiation blocking and transmitting areas. This corresponds to knife-edge measurements, which are a valid method to determine the spot size [43] [44] [45]. Their transition widths are determined by fitting error functions (lines) to the experimental data (dots) providing the FWHM diameter of the THz beams. The exact used function is $A = \frac{A_0}{2} \cdot \mathrm{erf}\left(\sqrt{2\ln(4)} \cdot \frac{x+x_0}{\mathrm{FWHM}} + A_{\mathrm{off}}\right)$, where $A_0$ is the amplitude difference between transmitted and blocked THz signal, $x_0$ is the slit position offset and $A_{\mathrm{off}}$ takes the background noise into account. The extracted FWHM values are for the 1.5 THz beam $5.8 \pm 1.1$ μm, for the 1.0 THz beam $6.1 \pm 1.0$ μm and for the 0.5 THz beam $5.9 \pm 1.3$ μm. Because of the large step size used for the imaging process, the extracted beam parameters are in the same magnitude, but are expected to be smaller, due to stronger laser beam focusing. For a final verification, additional one-dimensional knife-edge measurements at various positions, using the minimum step size of 200 nm confirm the spot size with a FWHM diameter close to and below 5 μm. Figure 2d) shows the extracted amplitude data for the corresponding positions as dots together with the fitted function by solid lines for 1 THz (red), 0.5 THz (black) and 1.5 THz (blue), respectively. The smallest FWHM in the plane of the Au layer is measured for 1.0 THz with $4.9 \pm 0.4$ μm. At 0.5 THz the FWHM increases to $5.1 \pm 0.5$ μm and at 1.5 THz to $5.0 \pm 0.5$ μm. The expected resolution limit for imaging in the far-field is in the range from 100 μm to 300 μm for radiation from 1 THz to 3 THz. Here the application of the near-field imaging technique improves the resolution by a factor of 20 to 60.

To support the experimental approach, the THz beam broadening during propagation in the intermediate distance between THz generator and imaged structure, and its propagation towards the detector are investigated by simulation. In the first step, we study the beam expansion on the $d_1 = 300$ nm path through the SiO$_2$ spacer layer, corresponding to the z-direction in the experiment. For this purpose, the wave equation derived from Maxwell equations is solved for a THz pulse propagating through the SiO$_2$ spacer layer, by iterating the time from $t = 0$ ps in steps of $\Delta t = 10$ as up to $t = 1$ ps after generation

$$\frac{\partial^2 \vec{E}}{\partial t^2} = c^2 \vec{\nabla}^2 \vec{E}.$$

Here, $\vec{E}$ is the THz beam electric field amplitude, $c = \frac{c_0}{n}$ is the phase velocity of light considering the medium refractive index $n$ and $\vec{\nabla}$ denotes the spacial wave propagation. The used parameters are, a THz pulse with a duration of 250 fs, a Gaussian beam spot with a FWHM = 4.7 μm and a refractive index for THz wavelengths in SiO$_2$ of $n = 2$ [46] [47].

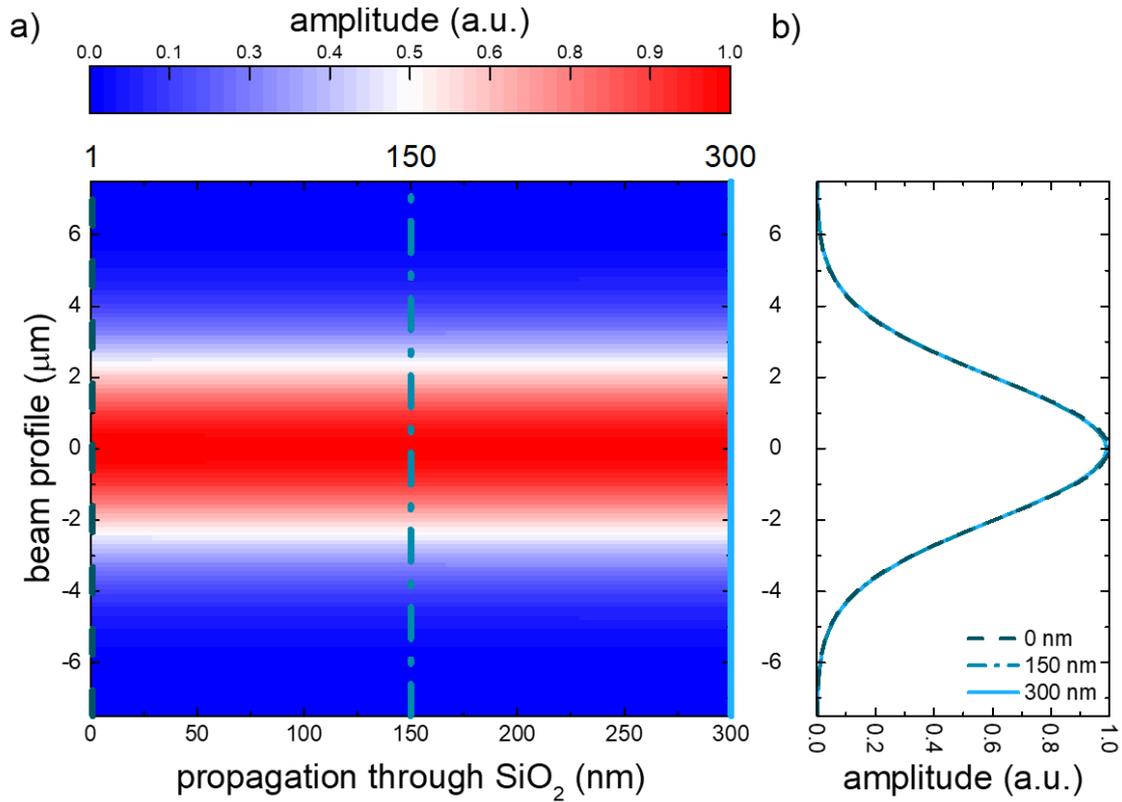

**Figure 3:** Simulation of the THz beam propagation through the **300 nm** thick SiO$_2$ layer. a) Snapshot of the THz beam e-field amplitude distribution **500 fs** after generation at **0 nm**. The Gaussian beam profile has a **FWHM = 4.7 μm** at the point of generation and broadens insignificantly after the propagation distance $d_1 = 300$ **nm**, as depicted in the extracted profiles in b).

The simulation results for the THz wave amplitude distribution within the SiO$_2$ spacer layer up to the imaged structure are illustrated in figure 3a), the snapshot is taken $500$ fs after emission from the Pt layer. The propagation through the SiO$_2$ clearly shows, that the beam width increases only insignificantly from the emission point to the imaging pattern. The trivial amplitude broadening at propagation distances, $d_1 = 150$ nm and $d_1 = 300$ nm, relative to the point of generation can be seen in figure 3b) indicating a very slight broadening below. Gauss fits to the amplitude distributions obtained throughout the whole time-interval reveal that the initial FWHM increases by around 1 % at this distance. This confirms, that defocusing happens on larger length scales and the THz beam stays focused within the spacer layer. Further simulations for larger distances even show, that propagation through spacer thicknesses in the micrometer range lead to a beam width expansion just below 10 %, if the refractive index of the spacer material stays below $n = 2$.

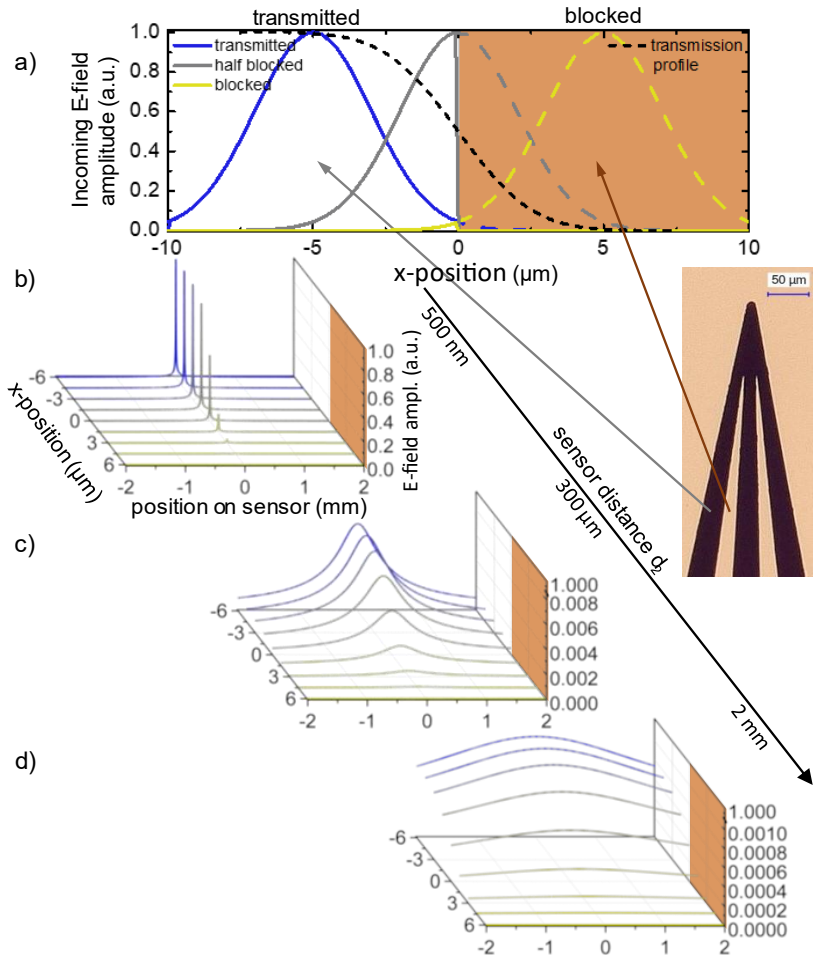

**Figure 4:** Intensity calculation for THz wave propagating the distance from slit to detector through air. a) Schematic depiction for three Gaussian beams passing the test pattern plane (orange) at three distances $d_2$ from the edge, showing different transmission scenarios (see text). At peak distances from edge $< 0$, the beam is fully (blue) or more than half (grey) transmitted, while at distances $\geq 0$, the beam is half (grey dashed) to entirely (yellow dashed) blocked. b-d) Calculated E-field amplitudes hitting the sensor after passing the test pattern plane depicted for three different detector distances $d_2$. At small distances around $d_2 = 500\,\text{nm}$ between test pattern and sensor, the arriving amplitude is high because the broadening of the beam is small. With increasing distance $d_2$, the THz wave broadens by three orders of magnitude, and the e-field amplitude decreases. For each sensor distance, the first 3 waves (x-position $< 0\,\text{mm}$) represented by blue colors of each distance are the largest, because they aren't cut off from the edge of the structure. The fifth wave (at x-position $= 0\,\text{mm}$) is half blocked by the structure, so that the amplitude is decreased to the half. The following waves (x-position $> 0\,\text{mm}$) are mostly or completely blocked, so that almost no or no more amplitude reaches the sensor.

The second simulation describes the behavior of the THz-beam after diffraction at the test pattern and propagating to the detector through air. This part of calculation is performed using energy conservation, Huygens–Fresnel principle and the superposition principle with complex-phase-corrected addition. Starting with a Gaussian electric field distribution in the test pattern plane, the electric field distribution arriving in the detector plane for different detector distances $d_2$ is calculated. The electric fields are normed to the amplitude in the test pattern plane at a distance of $d_1 = 300\,\text{nm}$ from the emitter after the propagation distance through the SiO$_2$. The computed results are imaged in figure 4.

Figure 4a) illustrates the beam passing the test pattern plane and leaving the sample. In this knife-edge scenario, the region with negative x-positions is fully transmitting (white) and for positive x-positions fully blocking the beam (orange). The knife-edge is located at x-position =

0 µm. Beam spots with a Gaussian electric field amplitude distribution whose center is in the transmitting region are fully or partially transmitted. Whereas beam spots whose center is located in the blocking region, are partially or fully blocked. Therefore, in this scenario, the transmitted beams are represented by the blue Gaussian distribution curve with the center position further away from the edge in the transmitting region than the beams FWHM. The beams partially blocked, with the center position closer to the knife-edge than the beams FWHM, are represented by the brown curve. The beams fully blocked, entering the pattern plane further away from the edge than their FWHM, are represented by the yellow curve. At this stage, the Gaussian beam has a FWHM = 4.7 µm. For the simulation, the Gaussian beam is moved from x-position $-10$ µm to $+10$ µm against the knife-edge and the diffraction pattern is calculated for an incrementally increasing sensor distance $d_2$. The resulting THz beam e-field amplitudes arriving at the sensor for each of the three sensor distances ($d_2 = 500$ nm, 300 µm, 2 mm) are depicted in figure 4b-d) respectively. For each sensor distance, the electric field amplitudes arriving at the sensor grouped in three groups, those fully transmitted are plotted in blue colors, those partially blocked are plotted in brown colors and those fully blocked are plotted in yellow colors. At the first sensor distance, 500 nm behind the test pattern, shown in figure 4b), only a small broadening occurs, and the peak electric field amplitude is in the range of the incoming wave. At larger detector distances in the order of the wavelength of 300 µm, the THz e-field distribution broadens into the millimeter scale (see figure 4c), and even further for the 2 mm sensor distance shown in figure 4d). At the same time, the THz peak electric field amplitude decreases with the sensor distance $\sim 1/d_2$, as known for spherical waves. Moreover, the amplitudes on the sensor decrease, when the x-position is changed from the transmitting to the blocking region across the knife-edge. In figures 4(b-d) this decrease is visible as a function of the x-position entering the test pattern plane. The width of this decrease corresponds to the width of the incoming beam set to $\text{FWHM} = 4.7$ µm, as confirmed by error-function fits.

Both simulation results, support the experimental data, showing, that the resolution of THz imaging with wavelengths $\lambda > 100$ µm, in the near-field approach can be enhanced to the micrometer range, limited only by the optical properties of the excitation light waves. Therefore, by using objectives for the pump laser beam, a resolution even below the micrometer scale will be possible. Besides this, the simulations show a very homogeneous and almost spherical wave propagation, which means, there are no interference minima that need to be avoided while adjusting the detector. Further, two parameters are revealed whose control will lead to a significant gain of measurement signal. First, the intensity of the THz wave propagating to the detector decreases by $\sim 1/d_2$ with the detector distance. This requires detector positioning within a vicinity of a few millimeters behind the emitter. Second, the intensity distribution broadens significantly due to the strong focusing of the exciting light beam for THz wave generation. Therefore, implementing collimating micro lenses directly into the plane of the imaged pattern would improve the signal. Lenses with such dimensions can be produced using three-dimensional lithography techniques directly on top of the emitters.

The results demonstrate a resolution enhancement for our approach applying near-field technique compared to the common far-field methods. For practical use, the spintronic THz emitters must be deposited directly on the planes or attached to the objects for imaging. This might be achieved by depositing THz emitters on glass surfaces of microscope slides or the corresponding coverslips. In addition, further improvement will be realized with the implementation of stronger focusing lenses and the application of other detector techniques to optimize the signal-to-noise ratio. However, the obtained results present the possibility to investigate significantly smaller structures by spectroscopy techniques in the future.

Our results clearly demonstrate the potential for near-field THz imaging using spintronic heterostructure based emitters. Their resolution can be enhanced below 1 µm and the emitter heterostructures can be deposited directly on glass surfaces e.g., microscope slides or coverslips. Adding nanometer thick spacer layers to those emitters, the distance to the imaged objects can be kept constant, ensuring identical conditions for THz wave generation on each spot and thus offering ideal conditions for imaging using the two-dimensional scanning technique. This provides the possibility for investigation of biological cells or e.g., blood platelets. In the future, more data revealing the spectral absorption of polymers in this range, will make this technique available for investigations of plastic nanoparticle residues in human cells. Also, insights into the specifics of spectral absorption will allow distinguishing between cell types and thus make the identification of e.g., cancer cells or blood platelet malformations possible. Besides this, this technique still offers signal-to-noise ratio improvement, by implementing collimating nanolenses to avoid beam divergence after the transmission through the imaged objects. This will not only increase signal intensity on the detector but also permit for more flexible detector positioning at larger distances to the samples. In addition, the implementation of other detection techniques, like electro-optic detection with ZnTe crystals will expand the investigated spectral range.

# Acknowledgements

The authors gratefully acknowledge the financial support from the BMBF, MetaZIK PlasMark-T (FKZ:03Z22C511).

Y.S. acknowledges the Graduate Program in Spintronics (GP-Spin) at Tohoku University.

# Data Availability

The data that support the findings of this study are available from the corresponding author upon reasonable request.